\documentclass[%
 reprint,
superscriptaddress,
nofootinbib,
 amsmath,amssymb,
 aps,
prd,
floatfix,
]{revtex4-2}
\usepackage{xcolor}
\usepackage{hyperref}
\usepackage{graphicx}
\usepackage{dcolumn}
\usepackage{bm}
\usepackage{tikz}
\usetikzlibrary{matrix,chains,positioning,decorations.pathreplacing,arrows}
\usepackage{float}
\usepackage[T1]{fontenc}
\usepackage{appendix}

\usepackage{lineno}

\begin{document}


\title{Fast and Flexible Analysis of Direct Dark Matter Search Data with Machine Learning}

\author{D.S.~Akerib} \affiliation{SLAC National Accelerator Laboratory, 2575 Sand Hill Road, Menlo Park, CA 94205, USA} \affiliation{Kavli Institute for Particle Astrophysics and Cosmology, Stanford University, 452 Lomita Mall, Stanford, CA 94309, USA} 
\author{S.~Alsum} \affiliation{University of Wisconsin-Madison, Department of Physics, 1150 University Ave., Madison, WI 53706, USA}  
\author{H.M.~Ara\'{u}jo} \affiliation{Imperial College London, High Energy Physics, Blackett Laboratory, London SW7 2BZ, United Kingdom}  
\author{X.~Bai} \affiliation{South Dakota School of Mines and Technology, 501 East St Joseph St., Rapid City, SD 57701, USA}  
\author{J.~Balajthy} \affiliation{University of California Davis, Department of Physics, One Shields Ave., Davis, CA 95616, USA}  
\author{J.~Bang} \affiliation{Brown University, Department of Physics, 182 Hope St., Providence, RI 02912, USA}  
\author{A.~Baxter} \affiliation{University of Liverpool, Department of Physics, Liverpool L69 7ZE, UK}  
\author{E.P.~Bernard} \affiliation{University of California Berkeley, Department of Physics, Berkeley, CA 94720, USA}  
\author{A.~Bernstein} \affiliation{Lawrence Livermore National Laboratory, 7000 East Ave., Livermore, CA 94551, USA}  
\author{T.P.~Biesiadzinski} \affiliation{SLAC National Accelerator Laboratory, 2575 Sand Hill Road, Menlo Park, CA 94205, USA} \affiliation{Kavli Institute for Particle Astrophysics and Cosmology, Stanford University, 452 Lomita Mall, Stanford, CA 94309, USA} 
\author{E.M.~Boulton} \affiliation{University of California Berkeley, Department of Physics, Berkeley, CA 94720, USA} \affiliation{Lawrence Berkeley National Laboratory, 1 Cyclotron Rd., Berkeley, CA 94720, USA} \affiliation{Yale University, Department of Physics, 217 Prospect St., New Haven, CT 06511, USA}
\author{B.~Boxer} \affiliation{University of Liverpool, Department of Physics, Liverpool L69 7ZE, UK}  
\author{P.~Br\'as} \affiliation{LIP-Coimbra, Department of Physics, University of Coimbra, Rua Larga, 3004-516 Coimbra, Portugal}  
\author{S.~Burdin} \affiliation{University of Liverpool, Department of Physics, Liverpool L69 7ZE, UK}  
\author{D.~Byram} \affiliation{University of South Dakota, Department of Physics, 414E Clark St., Vermillion, SD 57069, USA} \affiliation{South Dakota Science and Technology Authority, Sanford Underground Research Facility, Lead, SD 57754, USA} 
\author{N.~Carrara}  \email{ncarrara.physics@gmail.com} \affiliation{University at Albany, State University of New York, Department of Physics, 1400 Washington Ave., Albany, NY 12222, USA}   
\author{M.C.~Carmona-Benitez} \affiliation{Pennsylvania State University, Department of Physics, 104 Davey Lab, University Park, PA  16802-6300, USA}  
\author{C.~Chan} \affiliation{Brown University, Department of Physics, 182 Hope St., Providence, RI 02912, USA}  
\author{J.E.~Cutter} \affiliation{University of California Davis, Department of Physics, One Shields Ave., Davis, CA 95616, USA}  
\author{L.~de\,Viveiros}  \affiliation{Pennsylvania State University, Department of Physics, 104 Davey Lab, University Park, PA  16802-6300, USA}  
\author{E.~Druszkiewicz} \affiliation{University of Rochester, Department of Physics and Astronomy, Rochester, NY 14627, USA}  
\author{J.~Ernst} \affiliation{University at Albany, State University of New York, Department of Physics, 1400 Washington Ave., Albany, NY 12222, USA}  
\author{A.~Fan} \affiliation{SLAC National Accelerator Laboratory, 2575 Sand Hill Road, Menlo Park, CA 94205, USA} \affiliation{Kavli Institute for Particle Astrophysics and Cosmology, Stanford University, 452 Lomita Mall, Stanford, CA 94309, USA} 
\author{S.~Fiorucci} \affiliation{Lawrence Berkeley National Laboratory, 1 Cyclotron Rd., Berkeley, CA 94720, USA} \affiliation{Brown University, Department of Physics, 182 Hope St., Providence, RI 02912, USA} 
\author{R.J.~Gaitskell} \affiliation{Brown University, Department of Physics, 182 Hope St., Providence, RI 02912, USA}  
\author{C.~Ghag} \affiliation{Department of Physics and Astronomy, University College London, Gower Street, London WC1E 6BT, United Kingdom}  
\author{M.G.D.~Gilchriese} \affiliation{Lawrence Berkeley National Laboratory, 1 Cyclotron Rd., Berkeley, CA 94720, USA}  
\author{C.~Gwilliam} \affiliation{University of Liverpool, Department of Physics, Liverpool L69 7ZE, UK}  
\author{C.R.~Hall} \affiliation{University of Maryland, Department of Physics, College Park, MD 20742, USA}  
\author{S.J.~Haselschwardt} \affiliation{University of California Santa Barbara, Department of Physics, Santa Barbara, CA 93106, USA}  
\author{S.A.~Hertel} \affiliation{University of Massachusetts, Amherst Center for Fundamental Interactions and Department of Physics, Amherst, MA 01003-9337 USA} \affiliation{Lawrence Berkeley National Laboratory, 1 Cyclotron Rd., Berkeley, CA 94720, USA} 
\author{D.P.~Hogan} \affiliation{University of California Berkeley, Department of Physics, Berkeley, CA 94720, USA}  
\author{M.~Horn} \affiliation{South Dakota Science and Technology Authority, Sanford Underground Research Facility, Lead, SD 57754, USA} \affiliation{University of California Berkeley, Department of Physics, Berkeley, CA 94720, USA} 
\author{D.Q.~Huang} \affiliation{Brown University, Department of Physics, 182 Hope St., Providence, RI 02912, USA}  
\author{C.M.~Ignarra} \affiliation{SLAC National Accelerator Laboratory, 2575 Sand Hill Road, Menlo Park, CA 94205, USA} \affiliation{Kavli Institute for Particle Astrophysics and Cosmology, Stanford University, 452 Lomita Mall, Stanford, CA 94309, USA} 
\author{R.G.~Jacobsen} \affiliation{University of California Berkeley, Department of Physics, Berkeley, CA 94720, USA}  
\author{O.~Jahangir} \affiliation{Department of Physics and Astronomy, University College London, Gower Street, London WC1E 6BT, United Kingdom}  
\author{W.~Ji} \affiliation{SLAC National Accelerator Laboratory, 2575 Sand Hill Road, Menlo Park, CA 94205, USA} \affiliation{Kavli Institute for Particle Astrophysics and Cosmology, Stanford University, 452 Lomita Mall, Stanford, CA 94309, USA} 
\author{K.~Kamdin} \affiliation{University of California Berkeley, Department of Physics, Berkeley, CA 94720, USA} \affiliation{Lawrence Berkeley National Laboratory, 1 Cyclotron Rd., Berkeley, CA 94720, USA} 
\author{K.~Kazkaz} \affiliation{Lawrence Livermore National Laboratory, 7000 East Ave., Livermore, CA 94551, USA}  
\author{D.~Khaitan} \affiliation{University of Rochester, Department of Physics and Astronomy, Rochester, NY 14627, USA}  
\author{E.V.~Korolkova} \affiliation{University of Sheffield, Department of Physics and Astronomy, Sheffield, S3 7RH, United Kingdom} 
\author{S.~Kravitz} \email{swkravitz@lbl.gov} \affiliation{Lawrence Berkeley National Laboratory, 1 Cyclotron Rd., Berkeley, CA 94720, USA} 
\author{V.A.~Kudryavtsev} \affiliation{University of Sheffield, Department of Physics and Astronomy, Sheffield, S3 7RH, United Kingdom}  
\author{E.~Leason} \affiliation{SUPA, School of Physics and Astronomy, University of Edinburgh, Edinburgh EH9 3FD, United Kingdom}  
\author{B.G.~Lenardo} \affiliation{University of California Davis, Department of Physics, One Shields Ave., Davis, CA 95616, USA} \affiliation{Lawrence Livermore National Laboratory, 7000 East Ave., Livermore, CA 94551, USA} 
\author{K.T.~Lesko} \affiliation{Lawrence Berkeley National Laboratory, 1 Cyclotron Rd., Berkeley, CA 94720, USA}  
\author{J.~Liao} \affiliation{Brown University, Department of Physics, 182 Hope St., Providence, RI 02912, USA}  
\author{J.~Lin} \affiliation{University of California Berkeley, Department of Physics, Berkeley, CA 94720, USA}  
\author{A.~Lindote} \affiliation{LIP-Coimbra, Department of Physics, University of Coimbra, Rua Larga, 3004-516 Coimbra, Portugal}  
\author{M.I.~Lopes} \affiliation{LIP-Coimbra, Department of Physics, University of Coimbra, Rua Larga, 3004-516 Coimbra, Portugal}  
\author{A.~Manalaysay} \affiliation{Lawrence Berkeley National Laboratory, 1 Cyclotron Rd., Berkeley, CA 94720, USA} \affiliation{University of California Davis, Department of Physics, One Shields Ave., Davis, CA 95616, USA} 
\author{R.L.~Mannino} \affiliation{Texas A \& M University, Department of Physics, College Station, TX 77843, USA} \affiliation{University of Wisconsin-Madison, Department of Physics, 1150 University Ave., Madison, WI 53706, USA} 
\author{N.~Marangou} \affiliation{Imperial College London, High Energy Physics, Blackett Laboratory, London SW7 2BZ, United Kingdom}  
\author{D.N.~McKinsey} \affiliation{University of California Berkeley, Department of Physics, Berkeley, CA 94720, USA} \affiliation{Lawrence Berkeley National Laboratory, 1 Cyclotron Rd., Berkeley, CA 94720, USA} 
\author{D.-M.~Mei} \affiliation{University of South Dakota, Department of Physics, 414E Clark St., Vermillion, SD 57069, USA}  
\author{J.A.~Morad} \affiliation{University of California Davis, Department of Physics, One Shields Ave., Davis, CA 95616, USA}  
\author{A.St.J.~Murphy} \affiliation{SUPA, School of Physics and Astronomy, University of Edinburgh, Edinburgh EH9 3FD, United Kingdom}  
\author{A.~Naylor} \affiliation{University of Sheffield, Department of Physics and Astronomy, Sheffield, S3 7RH, United Kingdom}  
\author{C.~Nehrkorn} \affiliation{University of California Santa Barbara, Department of Physics, Santa Barbara, CA 93106, USA}  
\author{H.N.~Nelson} \affiliation{University of California Santa Barbara, Department of Physics, Santa Barbara, CA 93106, USA}  
\author{F.~Neves} \affiliation{LIP-Coimbra, Department of Physics, University of Coimbra, Rua Larga, 3004-516 Coimbra, Portugal}  
\author{A.~Nilima} \affiliation{SUPA, School of Physics and Astronomy, University of Edinburgh, Edinburgh EH9 3FD, United Kingdom}  
\author{K.C.~Oliver-Mallory} \affiliation{Imperial College London, High Energy Physics, Blackett Laboratory, London SW7 2BZ, United Kingdom} \affiliation{University of California Berkeley, Department of Physics, Berkeley, CA 94720, USA} \affiliation{Lawrence Berkeley National Laboratory, 1 Cyclotron Rd., Berkeley, CA 94720, USA}
\author{K.J.~Palladino} \affiliation{University of Oxford, Department of Physics, Oxford OX1 3RH, United Kingdon} \affiliation{University of Wisconsin-Madison, Department of Physics, 1150 University Ave., Madison, WI 53706, USA} 
\author{C.~Rhyne} \affiliation{Brown University, Department of Physics, 182 Hope St., Providence, RI 02912, USA}  
\author{Q.~Riffard} \affiliation{University of California Berkeley, Department of Physics, Berkeley, CA 94720, USA} \affiliation{Lawrence Berkeley National Laboratory, 1 Cyclotron Rd., Berkeley, CA 94720, USA} 
\author{G.R.C.~Rischbieter} \affiliation{University at Albany, State University of New York, Department of Physics, 1400 Washington Ave., Albany, NY 12222, USA}  
\author{P.~Rossiter} \affiliation{University of Sheffield, Department of Physics and Astronomy, Sheffield, S3 7RH, United Kingdom}  
\author{S.~Shaw} \affiliation{University of California Santa Barbara, Department of Physics, Santa Barbara, CA 93106, USA} \affiliation{Department of Physics and Astronomy, University College London, Gower Street, London WC1E 6BT, United Kingdom} 
\author{T.A.~Shutt} \affiliation{SLAC National Accelerator Laboratory, 2575 Sand Hill Road, Menlo Park, CA 94205, USA} \affiliation{Kavli Institute for Particle Astrophysics and Cosmology, Stanford University, 452 Lomita Mall, Stanford, CA 94309, USA} 
\author{C.~Silva} \affiliation{LIP-Coimbra, Department of Physics, University of Coimbra, Rua Larga, 3004-516 Coimbra, Portugal}  
\author{M.~Solmaz} \affiliation{University of California Santa Barbara, Department of Physics, Santa Barbara, CA 93106, USA}  
\author{V.N.~Solovov} \affiliation{LIP-Coimbra, Department of Physics, University of Coimbra, Rua Larga, 3004-516 Coimbra, Portugal}  
\author{P.~Sorensen} \affiliation{Lawrence Berkeley National Laboratory, 1 Cyclotron Rd., Berkeley, CA 94720, USA}  
\author{T.J.~Sumner} \affiliation{Imperial College London, High Energy Physics, Blackett Laboratory, London SW7 2BZ, United Kingdom}  
\author{N.~Swanson} \affiliation{Brown University, Department of Physics, 182 Hope St., Providence, RI 02912, USA}  
\author{M.~Szydagis} \affiliation{University at Albany, State University of New York, Department of Physics, 1400 Washington Ave., Albany, NY 12222, USA}  
\author{D.J.~Taylor} \affiliation{South Dakota Science and Technology Authority, Sanford Underground Research Facility, Lead, SD 57754, USA}  
\author{R.~Taylor} \affiliation{Imperial College London, High Energy Physics, Blackett Laboratory, London SW7 2BZ, United Kingdom}  
\author{W.C.~Taylor} \affiliation{Brown University, Department of Physics, 182 Hope St., Providence, RI 02912, USA}  
\author{B.P.~Tennyson} \affiliation{Yale University, Department of Physics, 217 Prospect St., New Haven, CT 06511, USA}  
\author{P.A.~Terman} \affiliation{Texas A \& M University, Department of Physics, College Station, TX 77843, USA}  
\author{D.R.~Tiedt} \affiliation{University of Maryland, Department of Physics, College Park, MD 20742, USA}  
\author{W.H.~To} \affiliation{California State University Stanislaus, Department of Physics, 1 University Circle, Turlock, CA 95382, USA}  
\author{L.~Tvrznikova} \affiliation{University of California Berkeley, Department of Physics, Berkeley, CA 94720, USA} \affiliation{Lawrence Berkeley National Laboratory, 1 Cyclotron Rd., Berkeley, CA 94720, USA} \affiliation{Yale University, Department of Physics, 217 Prospect St., New Haven, CT 06511, USA}
\author{U.~Utku} \affiliation{Department of Physics and Astronomy, University College London, Gower Street, London WC1E 6BT, United Kingdom}  
\author{A.~Vacheret} \affiliation{Imperial College London, High Energy Physics, Blackett Laboratory, London SW7 2BZ, United Kingdom}  
\author{A.~Vaitkus} \affiliation{Brown University, Department of Physics, 182 Hope St., Providence, RI 02912, USA}  
\author{V.~Velan} \affiliation{University of California Berkeley, Department of Physics, Berkeley, CA 94720, USA}  
\author{R.C.~Webb} \affiliation{Texas A \& M University, Department of Physics, College Station, TX 77843, USA}  
\author{J.T.~White} \affiliation{Texas A \& M University, Department of Physics, College Station, TX 77843, USA}  
\author{T.J.~Whitis} \affiliation{SLAC National Accelerator Laboratory, 2575 Sand Hill Road, Menlo Park, CA 94205, USA} \affiliation{Kavli Institute for Particle Astrophysics and Cosmology, Stanford University, 452 Lomita Mall, Stanford, CA 94309, USA} 
\author{M.S.~Witherell} \affiliation{Lawrence Berkeley National Laboratory, 1 Cyclotron Rd., Berkeley, CA 94720, USA}  
\author{F.L.H.~Wolfs} \affiliation{University of Rochester, Department of Physics and Astronomy, Rochester, NY 14627, USA}  
\author{D.~Woodward} \affiliation{Pennsylvania State University, Department of Physics, 104 Davey Lab, University Park, PA  16802-6300, USA}  
\author{X.~Xian} \affiliation{Brown University, Department of Physics, 182 Hope St., Providence, RI 02912, USA}  
\author{J.~Xu} \affiliation{Lawrence Livermore National Laboratory, 7000 East Ave., Livermore, CA 94551, USA}  
\author{C.~Zhang} \affiliation{University of South Dakota, Department of Physics, 414E Clark St., Vermillion, SD 57069, USA}  
\collaboration{LUX Collaboration}

\date{\today}

\begin{abstract}
We present the results from combining machine learning with the profile likelihood fit procedure, using data from the Large Underground Xenon (LUX) dark matter experiment. This approach demonstrates reduction in computation time by a factor of 30 when compared with the previous approach, without loss of performance on real data. We establish its flexibility to capture non-linear correlations between variables (such as smearing in light and charge signals due to position variation) by achieving equal performance using pulse areas with and without position-corrections applied. Its efficiency and scalability furthermore enables searching for dark matter using additional variables without significant computational burden. We demonstrate this by including a light signal pulse shape variable alongside more traditional inputs, such as light and charge signal strengths. This technique can be exploited by future dark matter experiments to make use of additional information, reduce computational resources needed for signal searches and simulations, and make inclusion of physical nuisance parameters in fits tractable.

\end{abstract}

\maketitle

\section{Introduction}\label{intro}

Xenon-based time projection chambers, such as LUX~\cite{bib:LUX2012}, excel at directly searching for dark matter in the form of weakly interacting massive particles (WIMPs). However, as these experiments grow in size and sensitivity, analysis procedures have  become increasingly complex and time-consuming. In estimating events above backgrounds either for exclusion limits or for discovery contours, the profile likelihood ratio (PLR) is the statistical analysis method of choice for most direct detection collaborations~\cite{Cowan:2010js, bib:LUXCal} having taken over from older ``cut and count'' approaches that used strict rectilinear cuts. 

While more accurate and more powerful than earlier approaches, the PLR is far slower and scales poorly with the number of observables. This leads to limitations on which variables can be practically included in the analysis, either as observable dimensions or as nuisance parameters. The goal of this work is to combine machine learning (ML) with the PLR to remove some of these limitations on the PLR approach. The motivations of our approach were as follows:

\begin{itemize}
\item Fewer Monte Carlo simulation statistics needed: ML is more efficient at capturing information to create models in the form of probability density functions (PDFs) than using binned histograms in a high-dimensional space.   
\item Variable independence is not a required assumption: Typically, observable signal sizes and positions are fed into a 4-5 dimensional model, which is broken up into independent 2D and 3D spaces to make computations tractable.
\item Faster computation time: ML collapses multiple variables for signal/background discrimination into only one 1D discriminant, taking much of the work out of generating PDFs, setting up the likelihood function, and designing PLR code to handle all of them appropriately in multiple dimensions. This is particularly valuable when increased model complexity is required, such as in the LUX second science run~\cite{LUX-fields, bib:LUXResults} where the electric field, fiducial mass, and signal gains were both spatially and temporally changing.
\item Full use of information: This approach is highly scalable with the inclusion of extra observable variables. In addition, while not explicitly addressed in this initial paper, evaluation of systematic uncertainties through the variation of nuisance parameters in the fit is made more feasible due to the faster calculation and lower statistics requirements.
\end{itemize}

The rest of the paper is organized as follows: in Sec.~\ref{sec:detector}, the basic design and operation of the LUX detector is explained. Sec.~\ref{sec:method} describes the common technique applied here of training a neural network to reduce the number of inputs to the PLR, along with details of the training procedure, such as how to ensure all relevant information is preserved. Sec.~\ref{sec:results} applies this technique to four different case studies: establishing its capability to reproduce a simple prior result; quantifying its ability to speed up the analysis; demonstrating its ability to use raw variables without loss of performance; and incorporating an additional S1 prompt fraction variable to demonstrate its scalability with additional inputs.

\section{The LUX Experiment}\label{sec:detector}

The technique employed by LUX is the dual-phase time projection chamber (TPC)~\cite{bib:TPC,bib:TPC2}. An incoming particle interacts with liquid xenon to produce scintillation light and ionization electrons~\cite{bib:LUX2012}. The ratio of ionization to scintillation depends on particle type, energy deposition, and electric fields. The difference in ratio between nuclear recoils (NR) and electronic recoils (ER) is the primary means for discriminating signal from backgrounds such as gamma-rays or electrons~\cite{bib:angle,bib:NR}. WIMPs should lead exclusively to NR events~\cite{bib:lewin} but non-WIMP dark matter may not~\cite{bib:nonwimp}.

An electric field drifts liberated electrons to the gas, where higher fields extract and drift them to make their own UV scintillation. The primary liquid scintillation signal is called S1, the secondary in gas, S2. S1 is the combination of photons from initial atomic de-excitations, and those from ionization electrons being re-captured into excited states. Electrons which are not captured escape to make the S2.

Xe detectors all search for WIMPs in a fashion independent of a specific model (e.g. Supersymmetry) by looking for excess NR events above the background. Neutrons can mimic WIMPs, but fewer background neutrons are produced than gamma-rays, so background ER is of the highest concern~\cite{bib:LUXmuon}. Underground deployment, and very aggressive material cleaning and screening campaigns, respectively, reduce cosmic-ray and intrinsic backgrounds dramatically~\cite{bib:lesko,bib:Mei}. Because of the remaining external radioactivity from cavern walls, internal sources like U and Th within the photomultiplier tubes (PMTs)~\cite{bib:PMT,bib:PMT2}, and most importantly Rn contamination from the environment \cite{LZ-radioassay}, discrimination of backgrounds at the level of data analysis remains a key requirement. It is largely achieved thanks to the S1 and S2 discrimination power: ER exhibits larger S2/S1 at fixed S1 than NR~\cite{vvelanDisc}. This discrimination is complicated by decays at the radial edge of the TPC (``wall backgrounds'') where the S2 signal is degraded, leading to partial overlap with the NR region.

The LUX detector housed 122 PMTs, with the xenon volume approximately 50 cm across and 60 cm tall between top and bottom PMT arrays~\cite{luxrun3_2013,bib:LUX2012}. The innermost 100-150 kg of 370 kg total were used as the fiducial mass. The S1 analysis threshold was 2 photons detected (phd, or spikes) corrected for the position dependence of the S1 photon detection efficiency, and the S2 threshold was 150-200 phd (6 to 8 extracted electrons), resulting in a 50\% detection threshold at 3-4 keV(nr)~\cite{luxrun3re2016}. It was deployed at the Sanford Underground Research Facility (SURF) in Lead, South Dakota, former site of the Homestake gold mine, at a depth of 4850 feet (4300 meters of water equivalent). The Xe was housed in a low-background Ti cryostat, which was itself housed within a water tank that further reduced background. In the TPC, the drift field was 180 V/cm and extraction field 6.0 kV/cm in the first science run, but 50-400 V/cm and 8.0 kV/cm, respectively, in the second~\cite{bib:LUXResults}. To calibrate the ER response and thus better understand the backgrounds as well as position resolution and monitor detector stability, $^{83m}$Kr, CH$_3$T, and $^{14}$CH$_4$ were injected during calibration acquisitions~\cite{bib:kr83m,bib:49,C14,GR_ERmodel}. For NR calibrations (for emulating the response of LUX to WIMPs) a D-D (deuterium-deuterium fusion) external neutron source was utilized~\cite{bib:LUXCal}.

These calibrations were used to define functions which correct for the position dependence of S1 and S2 signals, to ensure an approximately uniform mean signal size throughout the detector given a fixed energy. Such corrections, and the overall scaling factors for S1 and S2 signals versus energy, are not known perfectly; as such, it is common to incorporate these scaling factors as nuisance parameters in the PLR fit.

\section{Method}\label{sec:method}

The analyses in Sec.~\ref{sec:results} all follow essentially the same approach. First, a neural network (NN) \cite{bishop, Hastie2009} is trained using simulated data to distinguish events from a given dark matter signal model from backgrounds using a small set of high-level variables, $\mathbf{x}$, such as position, S1 area, and S2 area. The quality of the training is evaluated using several performance metrics including mutual information (MI), described further in Sec.~\ref{sec:MI}, as evaluated on an independent set of simulated testing data. In the case of suboptimal training, adjustments to the training procedure were made as detailed in Sec.~\ref{sec:nnarch}. In particular, enforcing that the MI for the NN output matches that of the input variables ensures that the NN transformation is optimal (preserves all relevant information) and provides an absolute calibration for when to stop the training process.

The output of the NN, $f(\mathbf{x}) \in [0, 1]$, which gives a monotonic indication of how signal-like an event is, is then transformed so that the test data, composed of equal parts background and signal events, follows a uniform distribution $g(f(\mathbf{x})) \in [0, 1]$. This is achieved by generating the cumulative distribution function (CDF) at any value, $f$, as the fraction of test events, $f'$, with $f'<f$, and then assigning $g(f(\mathbf{x})) = \mathrm{CDF}(f(\mathbf{x}))$, where linear interpolation is used between the $f'$ that appear in the test set. The purpose of this uniform transformation is to spread out the distribution, which the NN tends to focus at 0 and 1, so that the binned PDFs used in the PLR calculation better preserve information for events at the extreme values; an equivalent approach would be to use the original $f$ space but non-uniform binning which uses finer bins near the edges of the distribution. Including this transformation led to a small but noticeable improvement in PLR results during initial testing.

Once the NN is trained and the uniform transformation function is determined, both are applied to the simulated backgrounds, simulated signal, and the search data. The PLR calculation then proceeds in the usual fashion, using binned versions of the simulated signal and background distributions in this one-dimensional output variable $g(f(\mathbf{x}))$ as the PDFs in its likelihood function. It is performed in RooStats~\cite{RooStats} using code that is functionally identical to the non-NN approach, aside from the reduced complexity of a single input. Because each signal model (e.g. each WIMP mass) is different, a separate network is trained in each case, each with its own $g(f(\mathbf{x}))$; further details on techniques to ensure a smooth transition between similar models and avoid unnecessary duplicate training are provided in Sec.~\ref{sec:nnarch}. This approach is equally applicable to dark matter signal discovery as to limit-setting in the absence of signal, as the PLR procedure itself is unchanged.

The handling of systematic errors in the PLR calculation is not meaningfully changed by the use of the NN function $f(\mathbf{x})$, which is a well-defined and deterministic (if not simple to write down) function of the inputs, and as such cannot introduce any additional systematic uncertainties. As relevant nuisance parameters $\theta$ are varied within their uncertainties $\theta \rightarrow \theta'$, the distribution of the inputs to the NN, $p(\mathbf{x}|\theta)$, is adjusted accordingly to $p(\mathbf{x}|\theta')$. These changes are simply propagated through the NN output, resulting in a shift to the distribution $p(f(\mathbf{x})|\theta')$ (and hence the PDFs used in the fit) and a corresponding variation in the likelihood.

\subsection{Neural Network Architecture and Training}\label{sec:nnarch}

\begin{figure}
  \includegraphics[width=\linewidth]{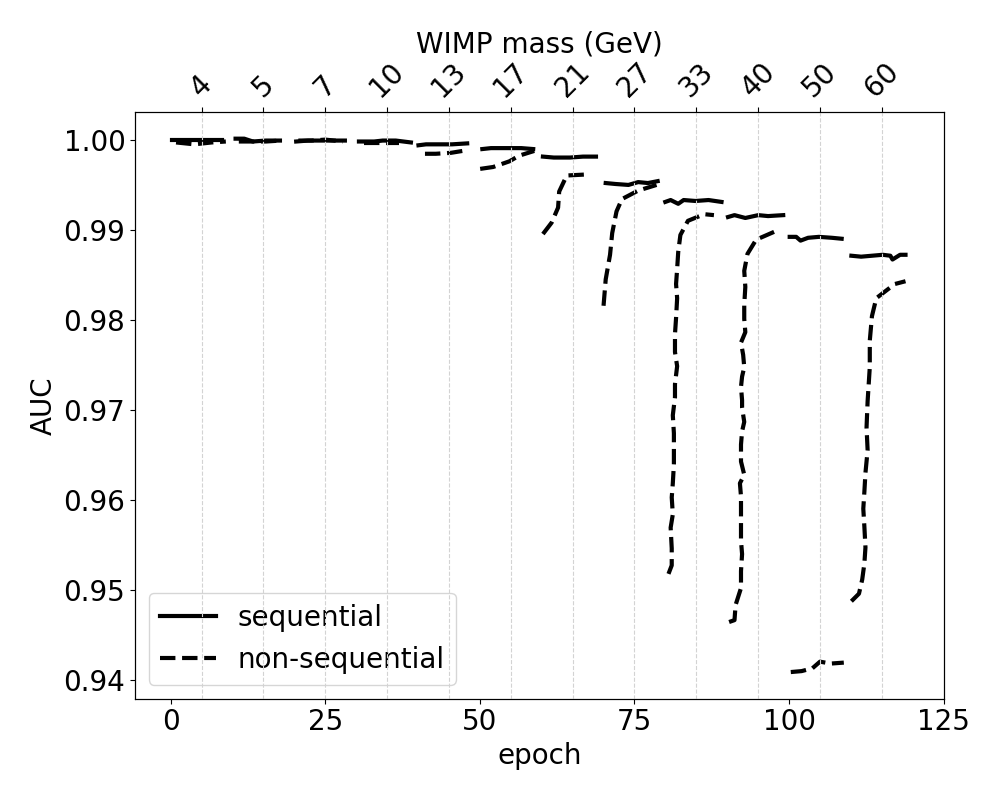}
  \caption{Training performance in Area Under Curve (AUC), which measures accuracy averaged across cut values, vs. training epoch. Every 10 epochs, a new WIMP mass is trained. The sequential method, using the network from the previous mass as the starting point for the next, converges more quickly and is less likely to find a suboptimal solution, as opposed to the non-sequential method (as in epochs 100-110). A subset of masses used during training is shown, to focus on relevant features.}
  \label{fig:transfer}
\end{figure}

Neural networks \cite{bishop, Hastie2009} in this work are implemented in the Python package Keras \cite{Keras}, which is a high level interface for Tensorflow \cite{tensorflow}. Sample code that illustrates the training procedure is available on GitHub\footnote{\href{https://github.com/luxdarkmatter/MachineLearningPublic}{https://github.com/luxdarkmatter/MachineLearningPublic}}. All networks use a simple sequential model, that is, a fully connected feed-forward network. Furthermore, all networks use a topology consisting of four layers which contain 4, 10, 3, and 1 nodes, respectively, with the exception of the work in Sec.~\ref{sec:psd}, which uses 5 nodes in the input layer due to the inclusion of an additional variable (S1 prompt fraction).

The training is conducted by using a sequential transfer learning technique in which we first train a network to distinguish a single WIMP mass from backgrounds, and then use the trained network parameters as the starting point to train on the next WIMP mass. Each training in this study begins with the smallest simulated WIMP mass of 3.5~GeV and proceed in sequence up to the largest. Fig.~\ref{fig:transfer} compares the training performance versus epoch (discrete training step) for sequential training against the case where a new network is trained with random initial weights at each mass. 

The main benefit of this technique is in the stability of training results across masses, due to the fact that the optimal network parameters should be similar for similar WIMP masses: poor local minima of the training optimization (as in epochs 100-110 of the non-sequential case) are unlikely to be found if the initial training is good. If the initial training is poor, this can be diagnosed quickly using the metrics explained below and retraining can occur, rather than having to retrain multiple faulty networks. An added benefit is the reduction of training times for each mass, since it takes fewer iterations to find optimum weights, as demonstrated by the relatively flat performance vs. epoch for the sequential case of Fig.~\ref{fig:transfer}. 

The trained NN performance is evaluated on an independent set of testing data in various metrics: the fraction of background events passing cuts at 50\% signal efficiency (``leakage''), the area under the signal efficiency vs. background leakage curve (AUC) -- a generalized version of leakage across all cut values in the NN output space, which varies from 0.5 (random guessing) to 1 (perfect discrimination) -- and mutual information (MI, Sec.~\ref{sec:MI}). If the networks at any mass showed signs of poor performance in these metrics, they were re-trained using a different set of initial weights (and in some cases a different number of training epochs or different batch size). This was rare, and hence re-training constituted only a minor increase in computation time. The low chance of requiring re-training is a good indication that the network optimization function is strongly correlated with these more physically-motivated metrics.

We note that traditional approaches at dimensional reduction such as principal component analysis (PCA) generally offer reduced complexity in the analysis space at the cost of reduced discrimination power, whereas this approach allows virtually no such loss as described in Sec.~\ref{sec:MI}. For a traditional method to achieve the same computational speed-up as this approach (Sec.~\ref{sec:timing}) requires reduction to a single dimension: the first principal component in the case of PCA. When applied to the data described in Sec.~\ref{sec:run3ws}, the background leakage at 50\% signal efficiency using the first principal component is greater than 10$\times$ worse than using a NN for intermediate masses where sensitivity is strongest, which would render a full analysis in the PCA-reduced space non-competitive.

\begin{figure}  
  \includegraphics[width=\linewidth]{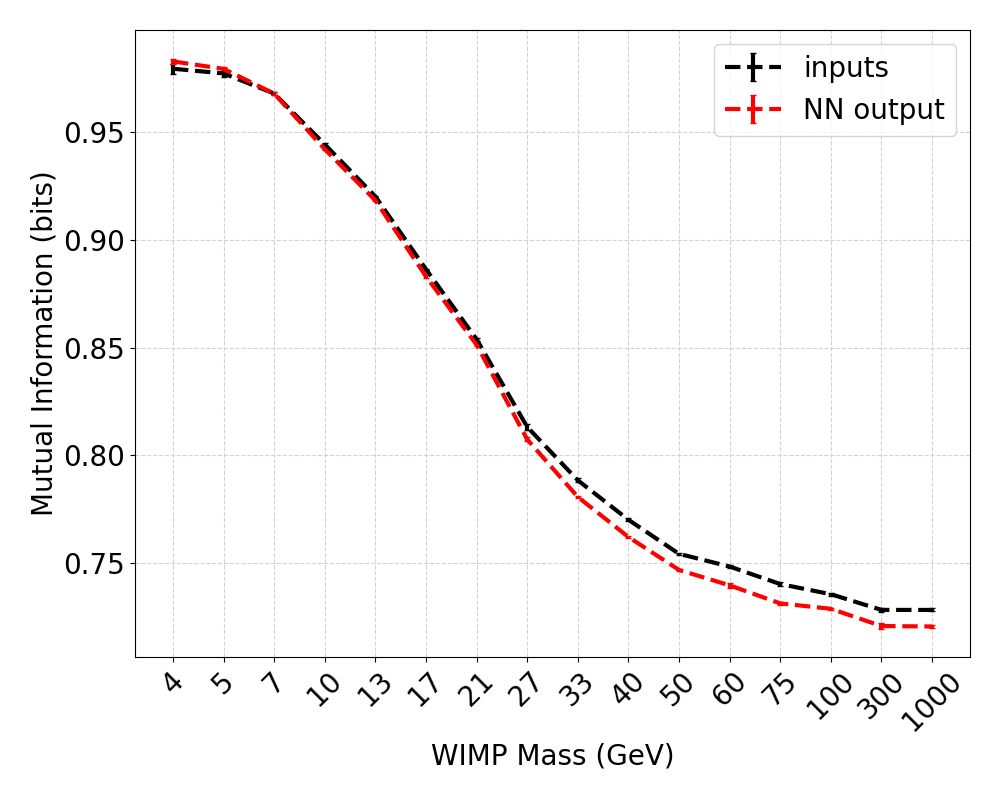}
  \caption{Mutual information (MI) between a set of discriminating variables -- the input variables ($S1,\log_{10}S2,r,z$) (black curve) or the trained NN output (red) -- and the classification of an event as signal or background, as a function of WIMP mass. The NN training achieves near-optimal results at all masses, indicating no loss of information from the reduction to a single dimension. The small discrepancy in MI at higher masses is negligible (see text for details). Errors on the MI estimate come from the statistical variation from re-calculating on multiple subsets, and are smaller than the line width at most masses. At the lowest two mass points, the MI from the NN is slightly higher than from the inputs; this is unphysical, but within the error on the MI estimate. }
  \label{fig:MI_vs_mass}
\end{figure}

\subsection{Mutual information}\label{sec:MI}

Apart from the standard metrics (AUC and leakage) we also adopt a technique which makes use of mutual information (MI) as a metric of performance for the NN training; specifically, it can be used to determine when the NN has equal power to discriminate signal from background as the full space of its inputs. The mutual information quantifies the strength of correlation between two sets of variables, $X$ and $Y$, which has the following functional form,
\begin{equation}
    I[x,y] = \int dxdy\, p(x,y)\log\frac{p(x,y)}{p(x)p(y)},
\end{equation}
where $p(x,y)$ is the probability distribution in $X$ and $Y$ (defined on a specific dataset), and $p(x)$, $p(y)$ are the marginal distributions, e.g. $p(x) = \int dy\, p(x,y)$.  

MI has been used in several different contexts from early work in information theory \cite{CoverThomas} to novel machine learning approaches \cite{Tishby}.  The context in which MI is used in this work was developed in \cite{CarraraVanslette,CarraraErnst}, where it was shown that the MI provides an \textit{upper-limit} on the performance of any machine learning algorithm. Specifically, whenever the MI between the binary signal/background class designation $\theta$ and the inputs $\mathbf{x}$ is equivalent to that between $\theta$ and the NN output $f(\mathbf{x})$, i.e $I[\theta;\mathbf{x}] = I[\theta;f(\mathbf{x})]$, the network is optimal: no information relevant to distinguishing signal from background events is lost in the transformation. 

Fig.~\ref{fig:MI_vs_mass} shows an example comparing the upper-limit, determined by computing the MI on the input variables ($S1,\log_{10}S2,r,z$), to the MI computed on the output of the trained neural network for each mass, using the case from Sec.~\ref{sec:run3ws}. Details on how the MI is estimated are given in \cite{CarraraErnst}. Lower WIMP masses show a larger MI (signal and background are more readily distinguished) largely due to better discrimination in (S1, S2) space at lower energy \cite{LZ-sensitivity, vvelanDisc}. This plot and other similar ones demonstrate that the NN has learned to summarize all relevant input information in a single dimension. The small discrepancy in MI at higher masses is negligible, corresponding in the worst case (1000~GeV) to a change in leakage for this analysis from 5.3e-3 to 5.8e-3. This is equivalent to a toy example of separating two 1D Gaussians of equal $\sigma$, where the separation of means decreases from 2.84 $\sigma$ to 2.81 $\sigma$.

We note that the MI estimate from this technique can be below the true value in a case with a large number of variables relative to available sample size in which some variables contribute no classification information. This limitation is not relevant for the 4D space considered here (Fig.~\ref{fig:MI_vs_mass}), chosen for its robust and well-studied modeling in simulation. MI can also be used to identify such variables from a large set of inputs by comparing the MI across different subsets of inputs, as in \cite{CarraraErnst}. This was observed, for example, in the analysis in Sec.~\ref{sec:psd} with the addition of an S1 prompt fraction variable. However, it is generally the case that modern NNs are capable of easily handling significantly more inputs than are used here, without loss of discrimination power (from uninformative variables) or prohibitive increases in required training time, such that tuning the set of inputs or significant adjustments to the training procedure are unnecessary.

\section{Results}\label{sec:results}
In Sec.~\ref{sec:run3ws}, we establish that the NN approach is capable of achieving equal WIMP detection limits to the traditional approach, using data from the first data-taking run of LUX. Sec.~\ref{sec:timing} demonstrates a speed-up from this technique of roughly 30$\times$ over the traditional approach, considering the more complicated structure of the second science run as applied to an effective field theory search. With the speed and efficacy of the approach established, we show in Sec.~\ref{sec:raws1s2} the network's flexibility to incorporate strongly-correlated variables by achieving equal limits for the first science run WIMP search using S1 and S2 areas before position-corrections are applied. To conclude, Sec.~\ref{sec:psd} incorporates an S1 prompt fraction variable into the analysis, verifying the ease of including new variables in this approach.

\begin{figure}
  \includegraphics[width=\linewidth]{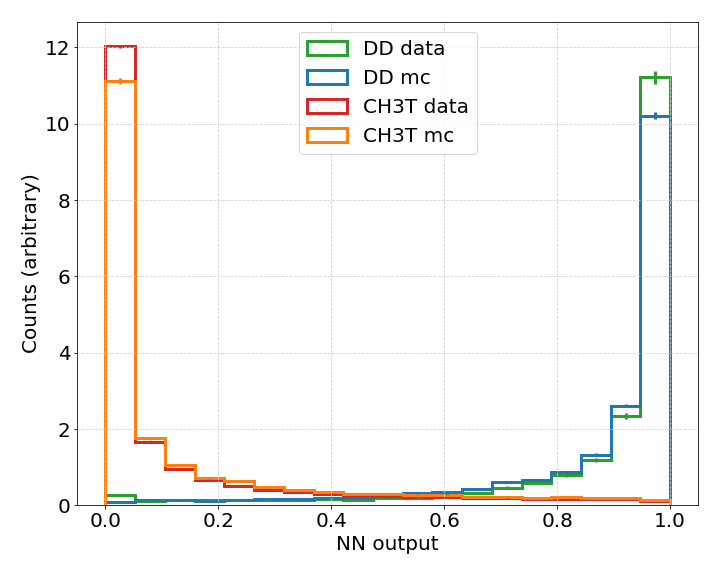}
  \caption{Monte Carlo simulation validation plots of NN output against ER calibration data (CH$_3$T) and NR calibration data (DD). Statistical error bars are shown; for most bins, they are comparable to the line width. As expected, the ER source appears similar to the training backgrounds (NN output closer to 0), while the NR source appears similar to the WIMP training signal (output near 1). The minor discrepancies between simulations and data at the extremes are of a similar size to that of the underlying inputs to the NN, and indicate the simulations are conservative: data for both the background-like CH$_3$T and signal-like DD events are more readily-identified as such, relative to the simulations.}
  \label{fig:NN_validation}
\end{figure}

\begin{figure*}
  \includegraphics[width=0.49\linewidth]{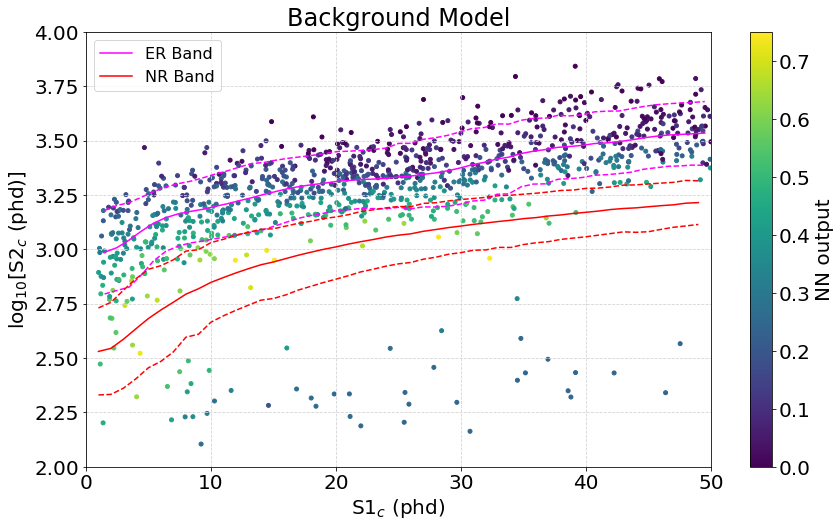}
  \includegraphics[width=0.49\linewidth]{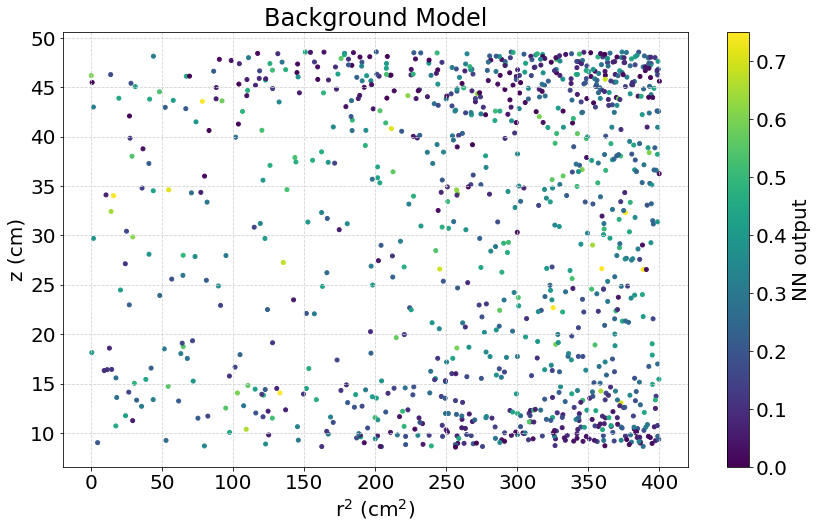}
  \includegraphics[width=0.49\linewidth]{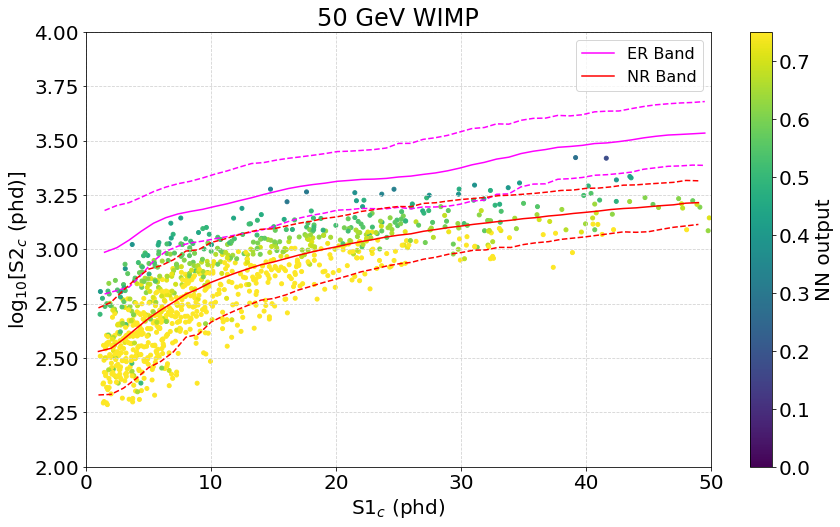}
  \includegraphics[width=0.49\linewidth]{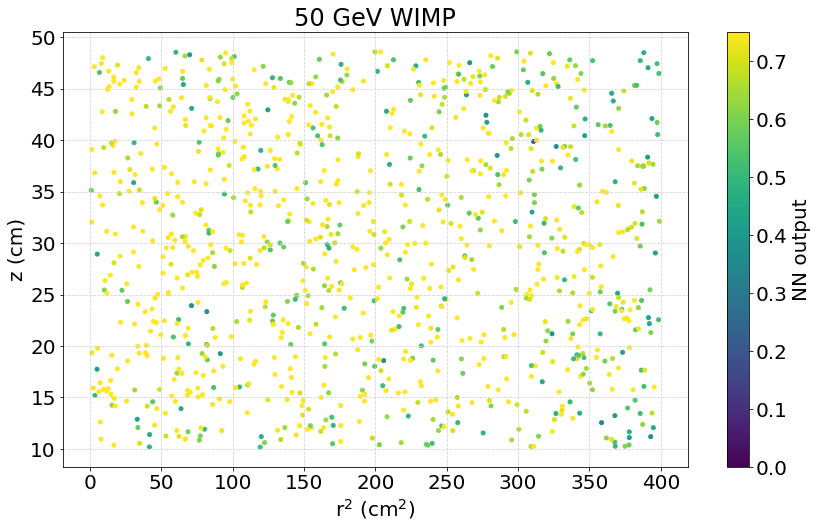}
  \includegraphics[width=0.49\linewidth]{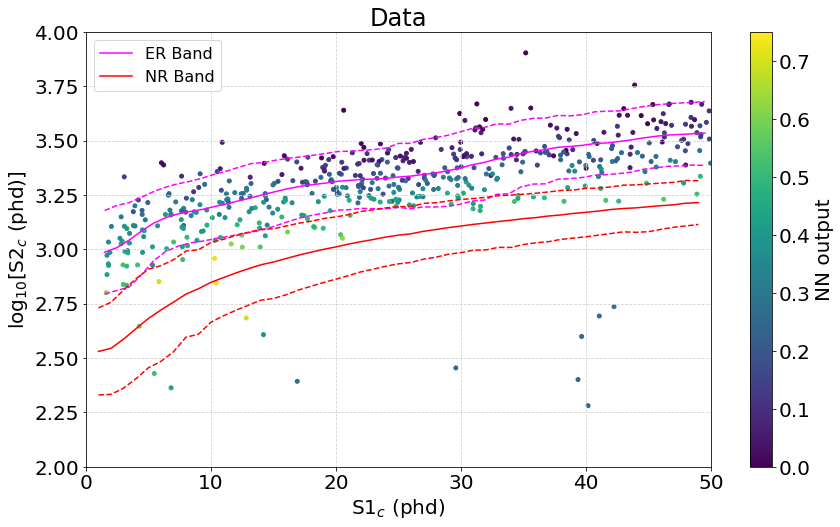}
  \includegraphics[width=0.49\linewidth]{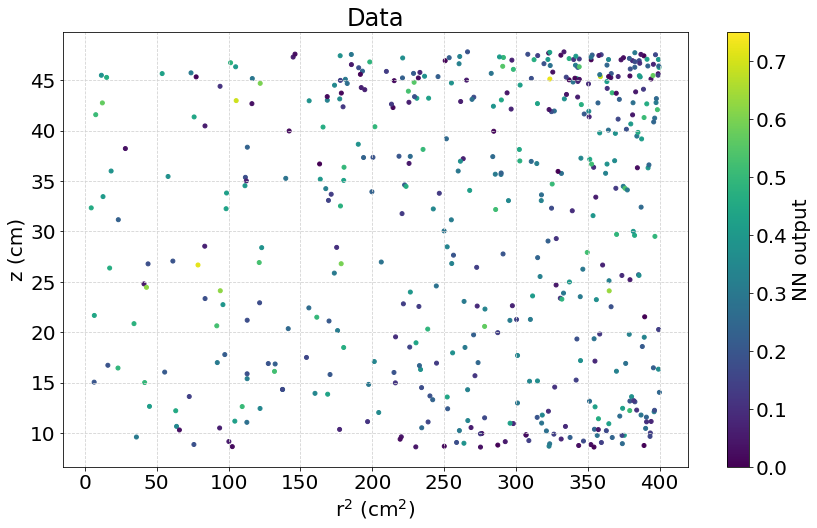}
  \caption{Visualized NN output for simulated background (top), 50 GeV WIMP signal (middle), and data (bottom) events in 2D input subspaces of $\log_{10}(S2_c)$ vs. $S1_c$ (left) and $z$ vs. $r^2$ (right), for the network trained against a 50 GeV WIMP in the 2013 reproduction analysis. Background and signal model visualizations use 1000 randomly-sampled events. Note that the color scale for the NN output is truncated at 0.7, roughly the maximum value observed in the data, to visually emphasize the small fraction of events in the data that are only slightly signal-like. Bands indicate the median, 10\%, and 90\% intervals.}
  \label{fig:input_vis}
\end{figure*}

\subsection{Reproduction of 2013 WIMP Search}\label{sec:run3ws}

This analysis uses data from the 2013 data-taking run of LUX \cite{luxrun3_2013}, with the intent of reproducing the results published in \cite{luxrun3re2016}. Analysis cuts, including data quality, fiducial volume (spatial cuts), and cuts on S1 and S2 range match those defined in \cite{luxrun3re2016}. The simulated background model is also the same as in \cite{luxrun3re2016}, and includes gamma-rays from multiple locations, beta particles, $^{127}\mathrm{Xe}$, $^{37}\mathrm{Ar}$, and backgrounds originating from decays at the radial edge of the TPC (``wall backgrounds''). WIMP dark matter signal models covering a range of masses are generated using the Noble Element Simulation Technique (NEST) version 2.0.1 \cite{NESTv2}, tuned to match LUX calibration data as described in \cite{LUX-DD, bib:LUXCal, GR_ERmodel}. 

The variables used as NN inputs were radial ($r$) and axial ($z$) position, position-corrected S1 area ($S1_c$), and the logarithm of the position-corrected S2 area ($\log_{10}(S2_c)$), as in \cite{luxrun3re2016}. However, whereas that analysis defined PDFs for most components of the form $f(r, z) \times f(S1_c, \log_{10}(S2_c))$, assuming independence of spatial and position-corrected pulse areas, the NN is able to account for relevant correlations in all variables. After sequential training of NNs at each WIMP mass as described in Sec.~\ref{sec:nnarch} and ensuring no loss of information from the NN transformation (Fig.~\ref{fig:MI_vs_mass}), the networks were validated against calibration data. Fig.~\ref{fig:NN_validation} shows the distributions of outputs of an example NN (trained against a 50~GeV mass WIMP, chosen for its proximity to the mass where the LUX sensitivity is maximized) for simulations and data for the standard ER and NR calibration sources used during the 2013 run. Both cases show good agreement, with a small excess at the background-like (signal-like) tails for the ER (NR) source in data, indicating that simulations are slightly conservative - the ER and NR calibration source distributions are more readily separated in data than in simulation.

An illustration of the information used by this sample 50~GeV WIMP NN is shown in Fig.~\ref{fig:input_vis}. These scatter plots indicate that the network has identified the NR band in the $S2$ vs. $S1$ space as signal, while ignoring as background wall events at the radial edges with low S2 values. 

\begin{figure}
  \includegraphics[width=\linewidth]{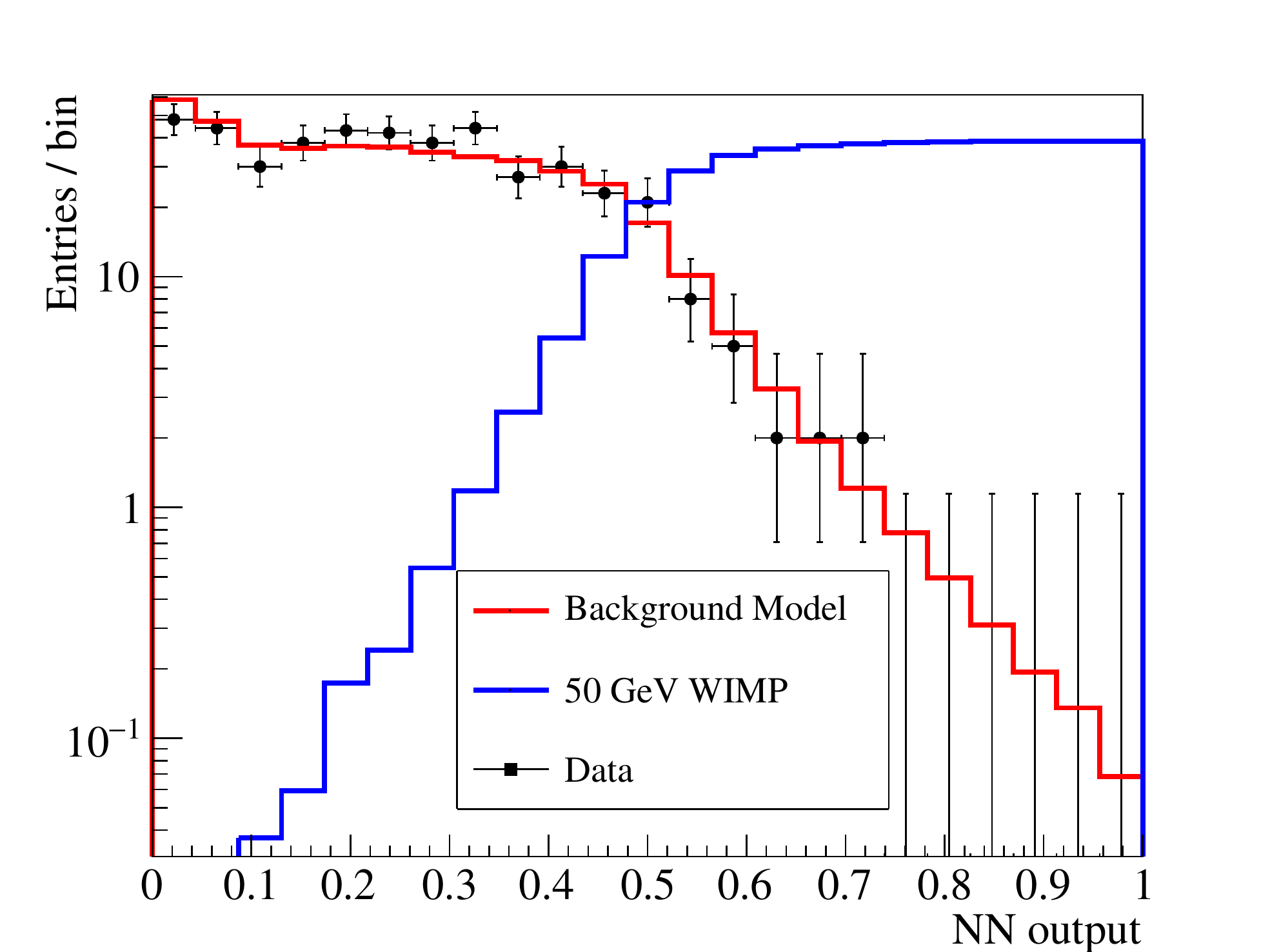}
  \caption{NN output space PDFs used in the fits for the 2013 WIMP search reproduction analysis, as well as binned data, for an example network trained against a 50 GeV WIMP. The output has been stretched using the uniform transformation described in Sec.~\ref{sec:method} to distinguish events at the extremes of the output range in this binned space. The data is consistent with a background-only hypothesis.}
  \label{fig:NNPDFs}
\end{figure}

The distribution of the search data in this 50~GeV WIMP NN output space is shown in Fig.~\ref{fig:NNPDFs}, along with the signal and background PDFs used in the fit. As expected from prior analyses using this data, it is consistent with a background-only hypothesis. 

The full PLR analysis procedure is then performed for each WIMP mass. As in the benchmark analysis \cite{luxrun3re2016}, the profile likelihood includes a Gaussian constraint for each background normalization nuisance parameter. However, in this analysis, all backgrounds are combined into a single component, with its constraint so broad ($\sigma=\mu=450$ events, equivalent to the total events in the search data) as to be inconsequential. This approach is maximally conservative, as it does not incorporate the information from standalone background studies to constrain the fit. Despite this, trials using multiple background components (each with their own PDF shape in the NN output space) and the stronger constraints enumerated in \cite{luxrun3re2016} showed no noticeable improvement in limit results (well within the $1\sigma$ confidence interval), indicating that the NN efficiently captures all relevant information without needing such standalone studies. 

\begin{figure}
  \includegraphics[width=0.99\linewidth]{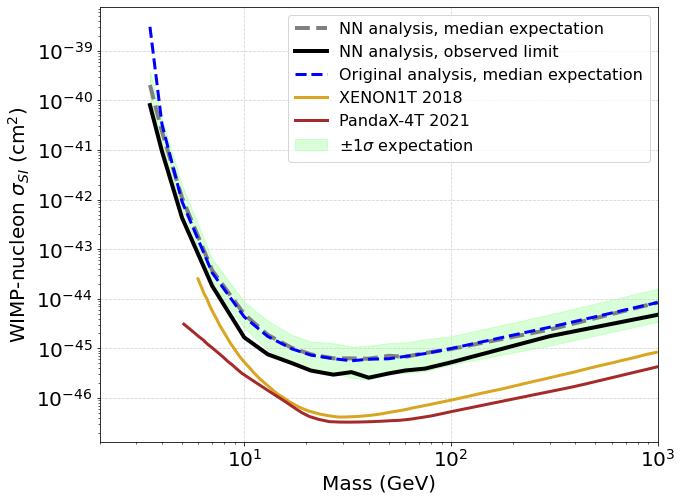}\hspace{0pt}
  \includegraphics[width=0.99\linewidth]{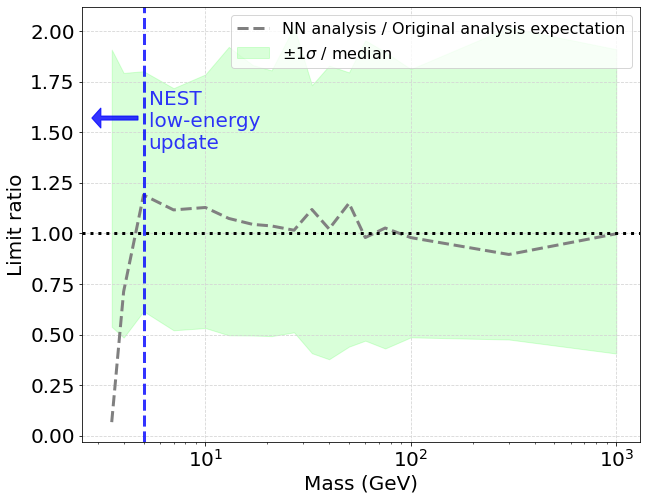}
  \caption{Top: Spin-independent WIMP-nucleon cross section 90\% CL upper limits for this work and the original analysis using the standard PLR approach on the same data from \cite{luxrun3re2016}. Limits from the much larger XENON-1T \cite{XENON1T2018} and PandaX-4T \cite{PandaX-4T} experiments are shown for context. This indicates equal performance when using the NN approach, both in simulations and in real data. Note that the original observed limit surpassed the median expectation, as in the NN case here, but was conservatively power constrained \cite{cowan-power-constraint} to the expectation. Bottom: ratio of limits using the NN approach presented here vs. the original limits. Values less than 1 indicate an improved limit using the NN analysis. Improvements below 5~GeV are primarily due to updates to the NEST yields model used to generate the signal events. Mass-to-mass fluctuations in expected limits of order 10\% are typical from prior non-ML analyses of LUX data \cite{LUX_EFT_Run4}.}
  \label{fig:2013limits}
\end{figure}

A summary of the limits produced by this procedure, relative to the \cite{luxrun3re2016} result, is given in Fig.~\ref{fig:2013limits}. This includes both the expected median sensitivity from simulations as well as the observed limit from the search data. These results indicate that the NN approach achieves virtually identical expected sensitivity (well within the $1\sigma$ confidence interval), with a small boost at masses below 5~GeV due to changes in NEST modeling at low energy. Notably, the observed limit outperforms the expectation from simulation, as it did in the original result (where it was power constrained \cite{cowan-power-constraint} to no better than the median expectation so as not to exclude cross sections for which sensitivity is low through chance background fluctuation), likely due to favorable statistical fluctuations and possibly also conservative modeling. This is a second confirmation, in addition to the validation on calibration data from Fig.~\ref{fig:NN_validation}, that the simulation-based NN training is reliably translated to real data.

\begin{table*}[t]
    \centering
    \caption{Comparison of total CPU hours required to generate a limit for a single signal model (e.g. single signal mass) for the standard PLR analysis described in \cite{LUX_EFT_Run4} vs. the NN approach, summed across all parallel processes. Workspace creation corresponds to building RooFit models containing the relevant PDFs and likelihood function; MC generation is the step of generating toy datasets from these PDFs - this is distinct from the full MC generation used to build the PDFs and train the NN; and hypothesis testing is the process of fitting the data and toys to the models and generating profile likelihood test statistic values. As some steps are not efficiently run in parallel, the speedup in real (wall-clock) time including all stages is further improved to 47$\times$. When run over a set of 24 masses for each of 15 EFT operators, the original CPU cost is approximately 18k CPU hours and the real time required to complete these computations is several days, making the addition of further complexity, such as nuisance parameters or additional analysis variables, impractical.}

    \begin{tabular}{l  c  c  c  c  c } 
        \hline \hline
         Analysis & Workspace creation (hr) & Toy MC generation (hr) & Hypothesis testing (hr) & NN training (hr) & \textbf{Total (hr)}\\
        \colrule
        Original EFT        & 2.7    & 39.0   & 8.8  & --   & \textbf{50.5} \\
        NN case             & 2.4e-3 & 1.0e-2 & 0.81 & 0.64 & \textbf{1.46} \\
        \colrule
        NN speedup          & 1100$\times$   & 3900$\times$   & 11$\times$   & --   & \textbf{35$\times$} \\
          
          \hline \hline
    \end{tabular}
    \label{tab:PLR_times}
\end{table*}

\subsection{Computational Speed-up}\label{sec:timing} 

In addition to simplifying the PLR while maintaining sensitivity, this approach greatly reduces the computational burden -- by a factor of $>$30x in this section's study, designed solely to quantify this speedup in a realistic limit-setting context. Because NNs are able to capture information in spaces with high dimensionality much more efficiently than histograms, the number of simulated events required for smooth PDFs is greatly reduced \cite{bishop, nn-density-estimation, Cranmer:2019eaq}. Additionally, the PLR limit-setting process itself is greatly sped up when run in a 1D space. This speed up remains substantial even after accounting for the time needed for NN training, as we demonstrate with the following comparison.

We take as our baseline the LUX analysis on effective field theory (EFT) couplings from its second science run \cite{LUX_EFT_Run4}. This analysis exemplifies the potential complexity of the standard PLR approach: the PDFs for most components are of the form $f(r, z) \times f(S1_c, \log_{10}(S2_c))$, i.e. the product of 3D and 2D PDFs, though the wall backgrounds are fully 5-dimensional. In addition, due to electric field variation over time and position, each PDF is broken up into four drift time bins and four date bins, for a total of sixteen PDFs per component. The fit includes eight background components, each with its own normalization nuisance parameter. Due to this complexity, even constructing the RooFit models necessary to hold these PDFs and generating the corresponding likelihood function can be time-intensive, taking nearly three hours per one mass value (see Tab.~\ref{tab:PLR_times}).

In contrast, the NN approach can directly incorporate position- and time-dependence in its structure, preserving the simplicity of the 1D output distribution. To allow for as direct a comparison as possible, the NN equivalent includes the same independent background components (and hence nuisance parameters allowed to vary in the fit). 

Tab.~\ref{tab:PLR_times} compares the compute time needed for the standard PLR approach as implemented in \cite{LUX_EFT_Run4} with the NN-based approach presented here. Accounting for all computational steps including the time required to train the NNs, Tab.~\ref{tab:PLR_times} demonstrates a reduction in computing time of 35$\times$ with the NN approach, and a reduction of 47$\times$ in the wall-clock time. This improvement factor is expected to increase with the number of inputs and the strength of their correlations, due to the NN's ability to account for complex relationships even in high dimensions without significant increase in computational demand; in contrast, the standard approach scales exponentially with further variables and requires additional, computationally-inefficient PLR structure to deal with correlations if independent PDFs are used (such as breaking up $f(S1_c, \log_{10}(S2_c))$ by drift time bin to preserve independence from $f(r,z)$). As the NN training comprises a significant fraction of the total compute time, this speedup may increase with more efficient training methods, such as the use of GPUs.

A full traditional PLR analysis including 15 EFT operators, each with 24 different masses, already stretches the limits of the computational resources that were available to LUX (approximately 18k CPU hours). The wall-clock time required for this style of analysis is an equally-important limitation, as a delay in results on the order of weeks is a significant setback, particularly given that iteration is typically required during analysis development before unblinding; in practice, a single iteration of the PLR analysis took several days and the full procedure, which required multiple iterations, required roughly two weeks of constant computation.  

This gain in processing speed makes it feasible to introduce further sophistication to the analysis through inclusion of PDF shape-varying nuisance parameters (such as linear scaling factors for S1 and S2 signals versus energy) in the fit, taking full advantage of strongly-correlated variables (Sec.~\ref{sec:raws1s2}), and even use of additional variables as inputs (Sec.~\ref{sec:psd}). From studies of this PLR analysis and similar traditional PLR approaches \cite{flamedisx}, we estimate the additional computational cost of increasing from four to five analysis dimensions to be 7-8$\times$, the cost of adding a single PDF shape-varying nuisance parameter to be 6-8$\times$, and the cost of increasing the number of observed events to be commensurate with those expected in the full 1000~day exposure of LZ to be 1.5$\times$. Extending the benchmark analysis from \cite{LUX_EFT_Run4} to include an additional dimension would thus take roughly a month for a single PLR iteration, and the actual workflow which required multiple iterations of this would have taken several months.

Other approaches to speed up the PLR procedure, notably Flamedisx \cite{flamedisx}, exist but lack the flexibility and modularity of this approach. Flamedisx uses an internal model of the detector and LXe microphysics to evaluate the likelihood of specific datapoints, requiring relevant detector effects for all variables of interest be implemented as analytic functions directly in the code. Events with multiple LXe interactions, such as some rare backgrounds, are also not modeled by Flamedisx. While this approach is more efficient than populating histograms in high-dimension, particularly when incorporating PDF shape-varying nuisance parameters, such modeling requirements limit its domain of application. In contrast, the NN approach is independent of the stages before and after it - any model (or combination of models) can be generated prior to the training, and a simple 1D PLR can be run after the training, allowing for checks at each stage.

\subsection{Use of Raw Variables}\label{sec:raws1s2}

This approach allows full use of the measured signal information, without assuming independence of position and pulse area variables, as in the traditional approach. Though position-correcting pulse areas improves sensitivity in the traditional case (through narrower ER and NR bands), even with perfectly-measured correction functions, the assumption of position-independence of pulse areas is not strictly accurate: corrections can adjust the mean of the distributions but not their widths, which vary due to position-dependence in light collection efficiency and electric field strength. In practice, this second-order effect of position-varying band widths is unlikely to significantly affect sensitivity unless there are strong nonuniformities in the detector, which was not the case for the 2013 LUX data considered here. However, this position correction approach is reliant on the assumption of well-measured position correction functions and sufficiently uniform detector response.

In contrast, the NN can account for correlations, making the step of applying position corrections to the pulse areas (and assuming position-independence) unnecessary. We demonstrate this by achieving equal sensitivity on LUX data whether using position-corrected pulse areas ($S1_c$, $S2_c$) or raw ones ($S1$, $S2$). This analysis uses the same background and signal models as in Sec.~\ref{sec:run3ws}, with the exception that the inverse position-correction functions are applied to the existing simulations to get $S1$ and $S2$, and a more sophisticated version of the wall background model is used, described in detail in \cite{LUX_EFT_Run4}. Notably, events near the walls are the most difficult to obtain reliable position corrections for, due to quickly-falling light collection efficiency with radius in that region \cite{bib:LUXCal, LUX-pos-recon}; this, coupled with the fact that these simulated events are generated in a data-driven way, makes use of the original uncorrected variables particularly well-suited to analysis of wall events.

\begin{figure}
  \includegraphics[width=\linewidth]{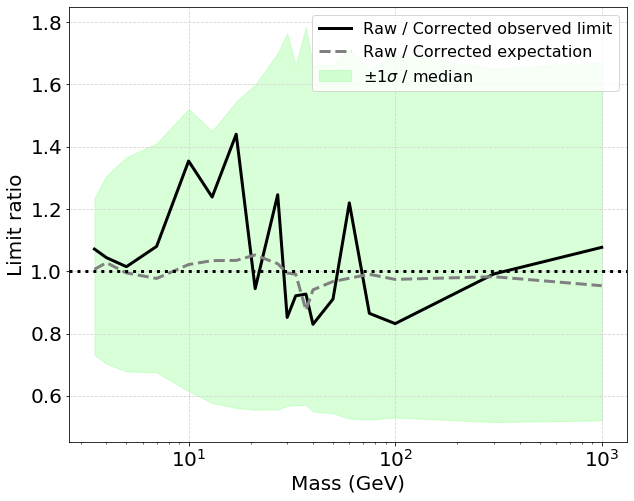}
  \caption{Ratio of limits using raw vs. position-corrected S1 and S2 pulse areas. Values less than 1 indicate an improved limit using the raw variables. Mass-to-mass fluctuations in expected limits of order 10\% are typical from prior non-ML analyses of LUX data \cite{LUX_EFT_Run4}; fluctuations in observed limits within the $1\sigma$ band are expected.}
  \label{fig:corr_raw_limits}
\end{figure}

The training procedure described in Sec.~\ref{sec:nnarch} was carried out for two cases, using identical NN architectures: one using inputs $\{r, z, S1_c, S2_c\}$ and another using $\{r, z, S1, S2\}$. Perhaps due to the additional complexity of the wall model events and/or the relative lack of wall events (limited, in this data-driven model, by the abundance of detected events in sidebands of the analysis region), some versions of the training in either variable space originally failed to identify the wall backgrounds reliably. This was easily spotted via the performance metrics described in Sec.~\ref{sec:method} as well as visualizations, such as those in Fig.~\ref{fig:input_vis}. Adjusting the training procedure, for each mass, to include a period of training with wall events as the sole background in between initial and final training with all backgrounds, solved this issue. 

After training was complete, no significant differences in performance on test simulation data were observed between the corrected and uncorrected versions of the networks. The full PLR procedure was then performed, this time allowing the wall and non-wall backgrounds to vary independently in the fits, with weak Gaussian constraints on their normalizations (25\% and 50\% for non-wall and wall backgrounds, respectively). A comparison of the limits achieved on the real data, as well as the model-based expected sensitivity, following the full PLR procedure is shown in Fig.~\ref{fig:corr_raw_limits}. Consistent with the training results, the expected and observed limits are comparable for the two cases.

While this approach does not directly eliminate the need for measurement of position-correction factors for pulse areas (these are needed for simulation-based models to produce raw detected pulse areas from true numbers of simulated photons and electrons), it avoids introducing potential errors in the estimate of these correction factors twice: once when simulating raw signals, and again when applying corrections to both simulations and data to get position-corrected variables. This second inverse correction does not undo the original mapping due to statistical fluctuations arising from the imperfect detection of generated signals. It is also a more natural way to analyze data-driven backgrounds, such as those from the walls, which need no such correction functions to determine raw signal size. Future work may be able to train directly on calibration data, learning such corrections implicitly without the need for extensive analyzer effort.

\subsection{Scalability with Increasing Number of Inputs}\label{sec:psd}

This approach is also easily scaled to include additional variables without significant computational burden or complexity. We demonstrate this through the use of an S1 prompt fraction variable, defined as in \cite{PulseShape}, and representing the fraction of the S1 pulse area within a fixed window at the start of the pulse. This variable carries information about the ER or NR nature of an event due to the differing ratios of singlet and triplet Xe excimers for the two recoil types, which in turn carry different photon emission times. Including this in the traditional PLR approach would be particularly challenging, as the S1 prompt fraction is correlated with both total S1 area and position (due to different photon propagation times to reach the PMTs), adding to the already-complex PDF structure, which otherwise assumes independence of position and pulse area variables. Such an analysis would be impractical to implement due to the computational resources required both to simulate a sufficient quantity of events to fill out binned PDFs in this expanded variable space and to carry out the PLR procedure: we estimate needing hundreds of thousands of CPU hours and months of real time with the traditional approach, as quantified in Sec.~\ref{sec:timing}.

Study of this S1 prompt fraction variable and its usefulness at ER/NR discrimination has been carried out in \cite{PulseShape, vvelanDisc}; however, validation of NEST-based simulations of the prompt fraction variable, including all correlations with pulse areas, position, and electric field strength, as necessary for this analysis, has not been performed until now. The details of this validation, using CH$_3$T and DD calibration data, are presented in Appendix~\ref{sec:psd_validation}.

\begin{figure}
  \includegraphics[width=\linewidth]{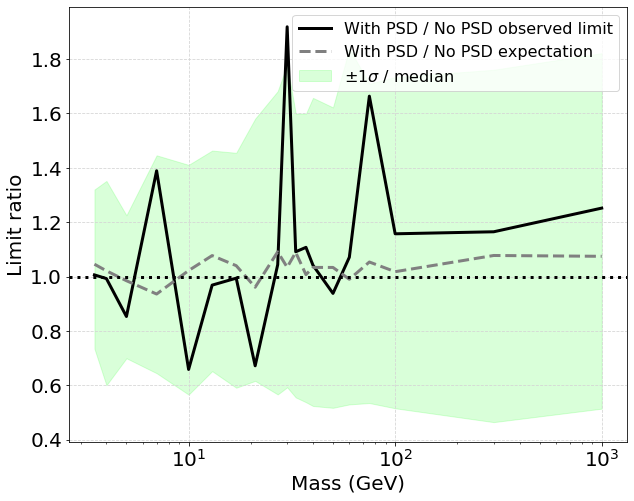}
  \caption{Ratio of limits using the S1 pulse shape discrimination variable vs. without it. Values less than 1 indicate an improved limit when including the PSD variable. Mass-to-mass fluctuations in expected limits of order 10\% are typical from prior non-ML analyses of LUX data \cite{LUX_EFT_Run4}; fluctuations in observed limits within the $1\sigma$ band are expected.}
  \label{fig:WS_psd_limits}
\end{figure}

For a more direct comparison with the analyses in Sec.~\ref{sec:run3ws} and \ref{sec:raws1s2}, the same background simulations were bootstrapped through NEST \cite{bootstrap-particle-physics}: the energies of the original simulated events were used to re-generate pulse areas and the new prompt fraction variable according to the NEST model, with multiple output events per input, sampling the random fluctuations in the model. Wall model events, being taken directly from sideband data, only require calculation of the prompt fraction using the observed waveform (including those corrections described in \cite{PulseShape}), as in the WIMP search data. The WIMP signal model events were generated directly in NEST using an identical model as for the previous analyses, but now including the validated prompt fraction variable.

Neural network training followed the same procedures as in previous sections, with the exception that the prompt fraction network had an additional input in its architecture (the size of the hidden layers was not adjusted). Inclusion of this variable had minimal impact on the training time of the networks. No significant differences in performance on test simulation data were observed between the two versions of the networks. This is consistent with MI calculations using the signal model and the non-wall backgrounds (wall events had insufficient statistics to get a reliable MI calculation), which indicate that while pulse shape has some relevant information on its own, it is mostly redundant when combined with other variables, such as S1 and S2 pulse area. This also matches expectations from \cite{vvelanDisc}, which suggest the ER leakage is not substantially improved by including pulse shape as part of a two-factor analysis (alongside S2/S1 ratio) in the 2013 LUX data, with its uniform 180~V/cm electric field.

The full PLR procedure was then performed, using the same Gaussian constraints as in Sec.~\ref{sec:raws1s2}; as the NN output is in a single dimension regardless of the number of inputs, the complexity and computation time of the PLR was not affected by adding the prompt fraction variable. As expected from the training stage, the limits achieved when including prompt fraction did not show an improvement over those without it (Fig.~\ref{fig:WS_psd_limits}), due to a lack of additional discriminating information in this context. Nevertheless, the flexible and scalable structure of this analysis framework made achieving this result straightforward. In a context where prompt fraction is expected to carry more information, such as in a scenario with lower electric field \cite{vvelanDisc, Kwong-psd, NEST-psd, XMASS-psd}, this approach could recover performance lost by the decreased ability to separate ER and NR sources using S1 and S2 area alone. Future work with next-generation experiments can easily extend the results here to include other variables of interest, such as topological discriminants relevant for multiple scatters as in \cite{EXO-deep} or track-like signal models.

\section{Conclusion}\label{sec:conclusion}
We have demonstrated a general-purpose approach to speed up and greatly improve the flexibility of dark matter direct detection limit-setting using machine learning. Its reliability is established both through checks against calibration data and its ability to reproduce the results of a prior dark matter search done with traditional methods. It achieves a speed-up of more than 30x the traditional method in a realistic test case, with stronger gains expected as more observables are considered. High sample size requirements from time-intensive simulations can likewise be reduced. In this way, it enables more complex analyses by ensuring their completion in days rather than months of real time.

This approach is flexible, with the capability to fully capture information from highly correlated variables, such as position and raw S1 and S2 pulse areas, without loss of performance. A future iteration of this may allow for the possibility of implicitly accounting for position corrections by training directly on calibration data, removing the need for analyzers to spend time manually defining analytic correction functions, as is required currently for accurate simulations. Furthermore, it is highly scalable in terms of computational demands, allowing the addition of further variables, such as S1 prompt fraction, without requiring any assumptions of variable independence. Future dark matter experiments can use this technique's efficiency to reduce their need for computation in both simulations and limit-setting, and its flexibility to enable analysis in a richer space, beyond that of pulse areas and positions.

\begin{acknowledgements}
This work was partially supported by the U.S. Department of Energy (DOE) under Awards No. DE-AC02-05CH11231, No. DE-AC05-06OR23100, No. DE-AC52-07NA27344, No. DE-FG01-91ER40618, No. DE-FG02-08ER41549, No. DE-FG02-11ER41738, No. DE-FG02-91ER40674, No. DE-FG02-91ER40688, No. DE-FG02-95ER40917, No. DE-NA0000979, No. DE-SC0006605, No. DE-SC0010010, No. DE-SC0015535, and No. DE-SC0019066; the U.S. National Science Foundation under Grants No. PHY-0750671, No. PHY-0801536, No. PHY-1003660, No. PHY-1004661, No. PHY-1102470, No. PHY-1312561, No. PHY-1347449, No. PHY-1505868, and No. PHY-1636738; the Research Corporation Grant No. RA0350; the Center for Ultra-low Background Experiments in the Dakotas (CUBED); and the South Dakota School of Mines and Technology (SDSMT).

Laborat\'{o}rio de Instrumenta\c{c}\~{a}o e F\'{i}sica Experimental de Part\'{i}culas (LIP)-Coimbra acknowledges funding from Funda\c{c}\~{a}o para a Ci\^{e}ncia e a Tecnologia (FCT) through the Project-Grant No. PTDC/FIS-NUC/1525/2014. Imperial College and Brown University thank the UK Royal Society for travel funds under the International Exchange Scheme (IE120804). The UK groups acknowledge institutional support from Imperial College London, University College London, the University of Sheffield, and Edinburgh University, and from the Science \& Technology Facilities Council for PhD studentships R504737 (EL), M126369B (NM), P006795 (AN), T93036D (RT), and N50449X (UU). This work was partially enabled by the University College London (UCL) Cosmoparticle Initiative. The University of Edinburgh is a charitable body, registered in Scotland, with Registration No. SC005336.

This research was conducted using computational resources and services at the Center for Computation and Visualization, Brown University, and also the Yale Science Research Software Core.

We gratefully acknowledge the logistical and technical support and the access to laboratory infrastructure provided to us by SURF and its personnel at Lead, South Dakota. SURF was developed by the South Dakota Science and Technology Authority, with an important philanthropic donation from T. Denny Sanford. SURF is a federally sponsored research facility under Award Number DE-SC0020216.
\end{acknowledgements}

\begin{appendix}
\section{Pulse Shape Discrimination Validation}\label{sec:psd_validation}

This appendix presents the validation of simulations of the S1 prompt fraction variable from Sec.~\ref{sec:psd} using calibration data. Similar validations were performed in \cite{PulseShape}, which we extend here by tuning the model to match the dependence on electric field and drift time seen in data, as well as to cover smaller S1 pulse areas, which account for the bulk of WIMP events. Such tuning is necessary, as correlations can in principle matter for the NN approach. This was done primarily through adjusting the Xe excimer singlet-to-triplet ratio and its dependence on energy and electric field \cite{Kwong-psd}. 

Adjustments to better account for detector-dependent effects were also included. Chief among these was the addition of a random offset in the prompt integration window, to account for the finite (10~ns) waveform sampling rate and other sources of timing uncertainty. When counting a few individual photons ($\lesssim$~10), the chance of getting the full S1 area within the prompt fraction window becomes substantial. To account for this, an empirical adjustment was applied, randomly assigning a fraction of events to a prompt fraction of 1 according to an falling exponential in S1 area, as fit to low-energy calibration data. 

At smaller still S1 areas, the model diverges from data in a non-trivial way. For S1 areas below 5~phd, we conservatively assign both simulations and data a prompt fraction drawn from a single Gaussian with a mean halfway between that of the ER and NR calibration sources at 5~phd and a standard deviation comparable to that of both, truncated to the allowable range of 0-1. This ensures the NN cannot learn any distinguishing features from the prompt fraction at these low areas (where it is of little use regardless due to the difficulty of defining a pulse shape from so few photons \cite{vvelanDisc}). The choice of Gaussian parameters was verified to have no noticeable effect on NN training.

\begin{figure*}[!ht]
  \includegraphics[width=0.49\linewidth]{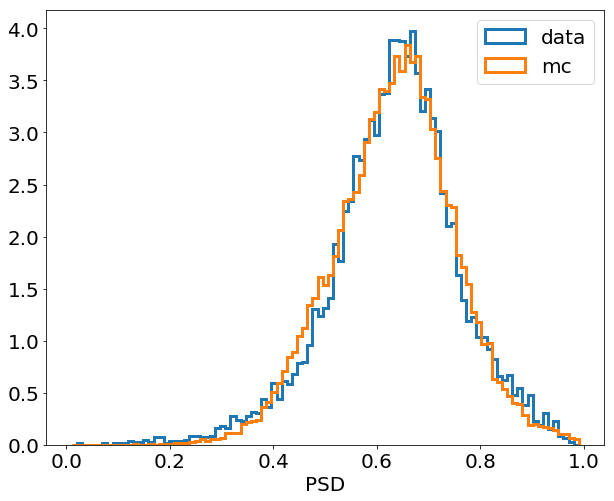}
  \includegraphics[width=0.49\linewidth]{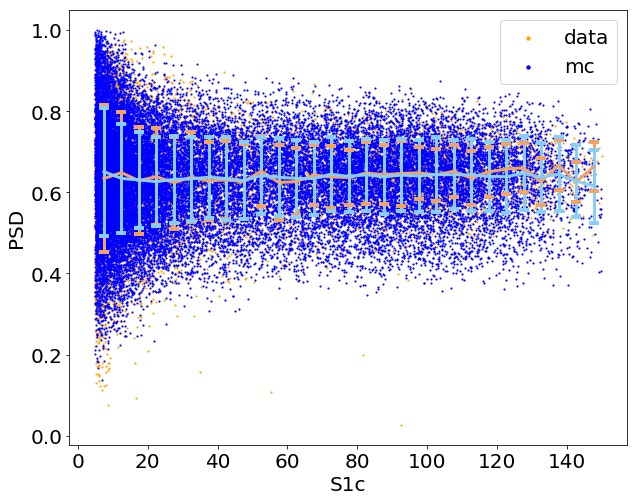}
  \includegraphics[width=0.49\linewidth]{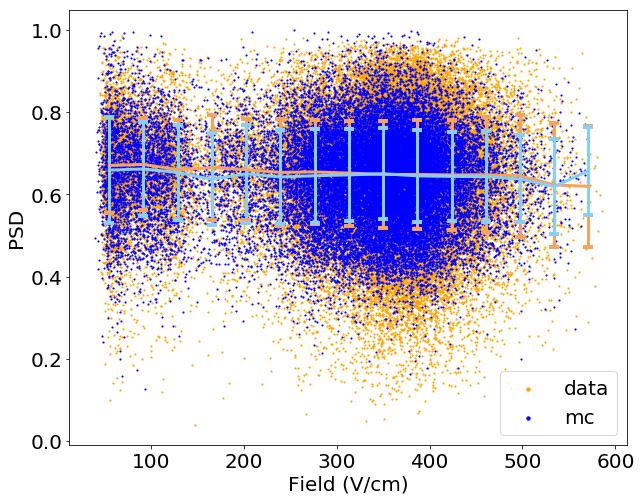}
  \includegraphics[width=0.49\linewidth]{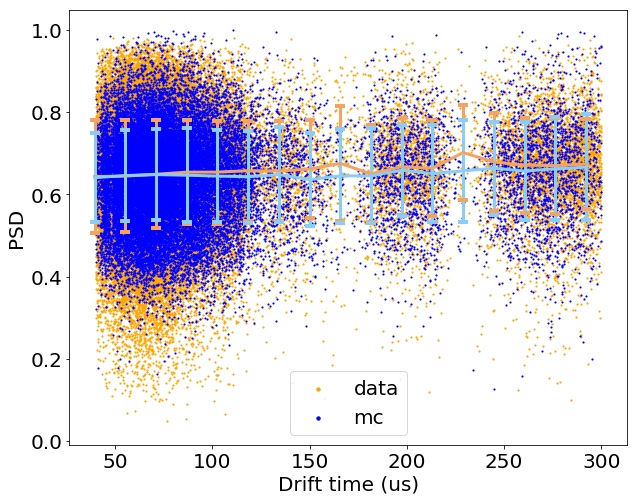}
  \caption{Comparison of MC simulations of S1 pulse shape discriminator (PSD) with that calculated from DD neutron calibration data. Lines indicate the mean PSD, with the standard deviation at each point indicated via error bars. Drift time and field only varied significantly in the 2014-2016 data-taking run, so calibration data from that run is used in the bottom two plots. The lobes in the bottom two plots correspond to the three different heights at which the DD generator was employed.}
  \label{fig:psd_DD_calibrations}
\end{figure*}

\begin{figure*}[!ht]
  \includegraphics[width=0.49\linewidth]{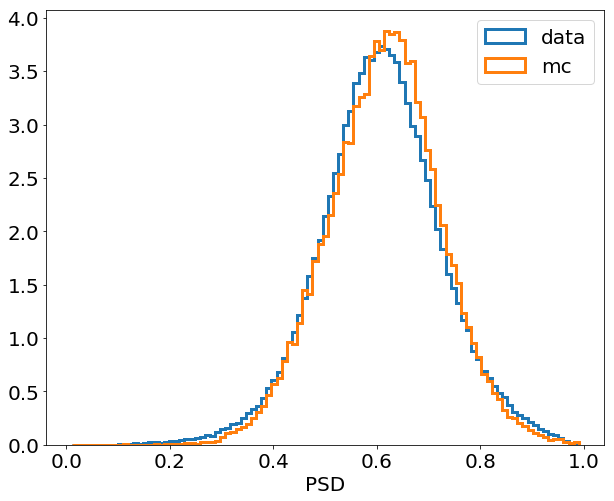}
  \includegraphics[width=0.49\linewidth]{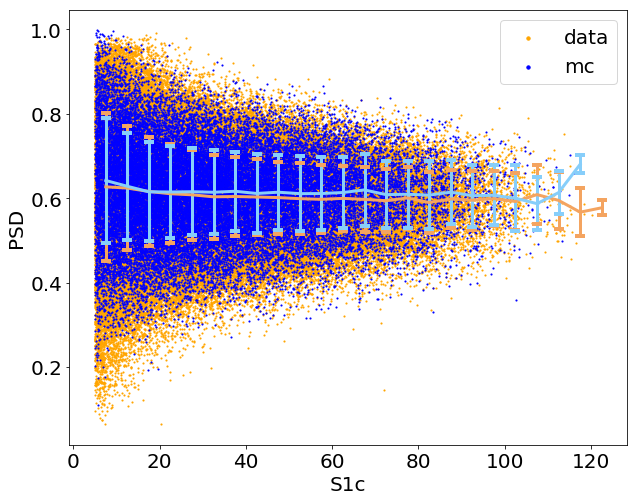}
  \includegraphics[width=0.49\linewidth]{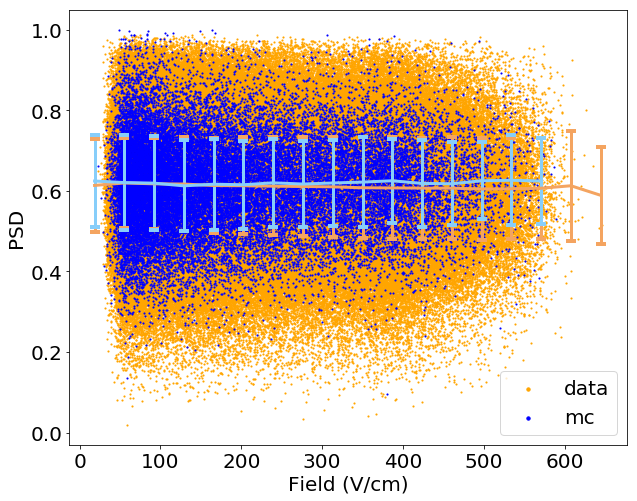}
  \includegraphics[width=0.49\linewidth]{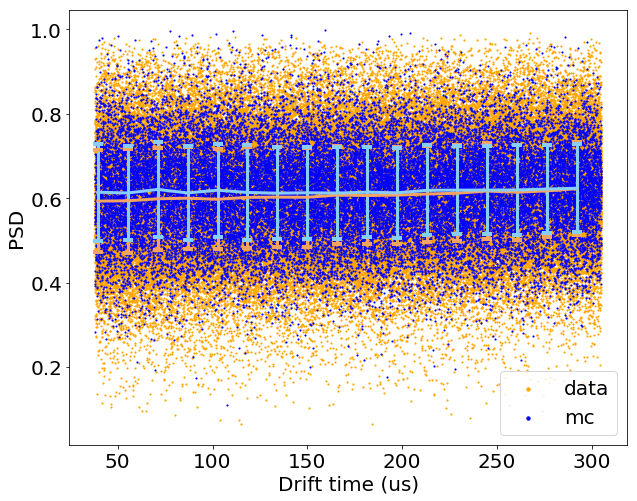}
  \caption{Comparison of MC simulations of S1 pulse shape discriminator (PSD) with that calculated from CH$_3$T ER calibration data. Lines indicate the mean PSD, with the standard deviation at each point indicated via error bars. Field only varied significantly in the 2014-2016 data-taking run, so calibration data from that run is used in the bottom left plot.}
  \label{fig:psd_CH3T_calibrations}
\end{figure*}

\begin{figure*}[!ht]
  \includegraphics[width=0.49\linewidth]{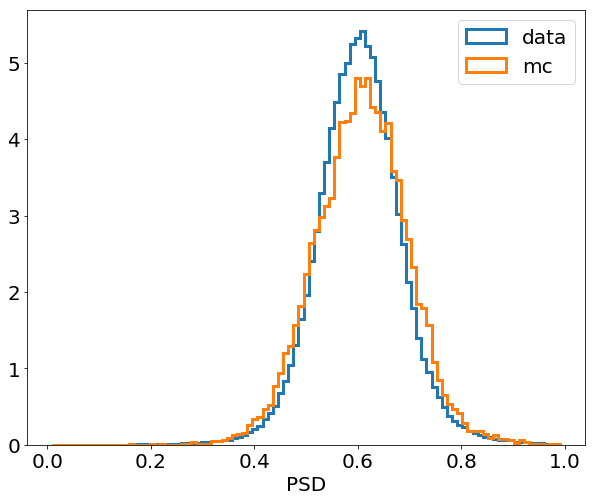}
  \includegraphics[width=0.49\linewidth]{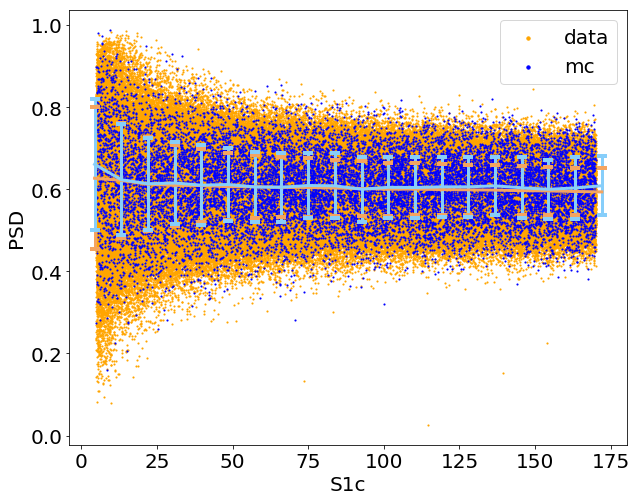}
  \includegraphics[width=0.49\linewidth]{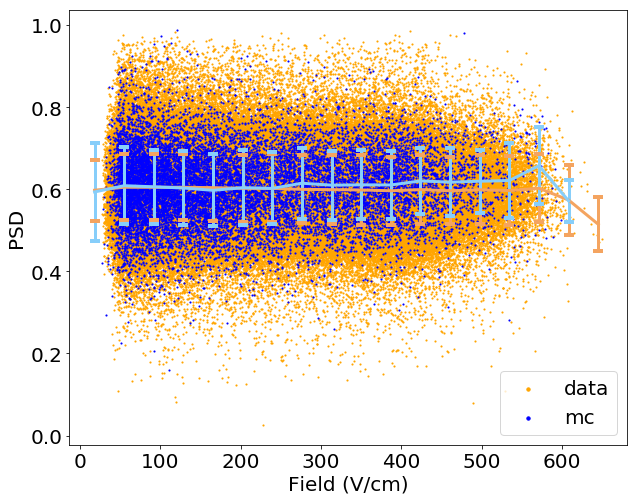}
  \includegraphics[width=0.49\linewidth]{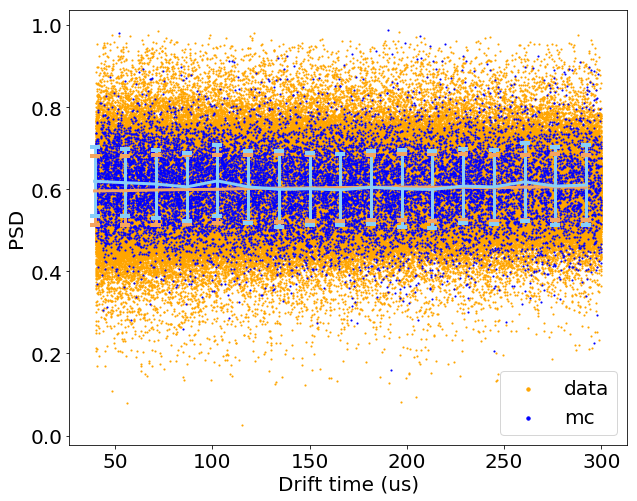}
  \caption{Comparison of MC simulations of S1 pulse shape discriminator (PSD) with that calculated from $^{14}$CH$_4$ high-energy ER calibration data. Lines indicate the mean PSD, with the standard deviation at each point indicated via error bars. All plots use conditions from the 2014-2016 data-taking run, as $^{14}$CH$_4$ calibration data was only taken during that run.}
  \label{fig:psd_C14_calibrations}
\end{figure*}

Validation plots of the pulse shape variable are shown for NR calibration with the DD neutron source (Fig.~\ref{fig:psd_DD_calibrations}), low-energy ER calibration with CH$_3$T (Fig.~\ref{fig:psd_CH3T_calibrations}), and high-energy ER calibration with $^{14}$CH$_4$ (Fig.~\ref{fig:psd_C14_calibrations}). The NN output for the model trained in Sec.~\ref{sec:psd}, as applied to the calibration sources, showed good agreement between simulations and data, similar to in Fig.~\ref{fig:NN_validation}, establishing that the full input space correlations learned by the NN are well-captured by the simulations.

\end{appendix}

\bibliography{main.bib}

\end{document}